# A Review of Internet of Things Architecture, Technologies and Analysis Smartphone-based Attacks Against 3D printers

**Muhammad Bilal**

Department of Computer Science
Zhejiang University Hangzhou, China
mbilal@zju.edu.cn

*Abstract*

*Human beings cannot be happy with any kind of tiredness based work, so they focused on machines to work on behalf of humans. The Internet-based latest technology provides the platforms for human beings to relax and unburden feeling. The Internet of Things (IoT) field efficiently helps human beings with smart decisions through Machine-to-Machine (M2M) communication all over the world. It has been difficult to ignore the importance of the IoT field with the new development of applications such as a smartphone in the present era. The IoT field sensor plays a vital role in sensing the intelligent object/things and making an intelligent decision after sensing the objects. The rapid development of new applications using smartphones in the world caused all users of the IoT community to be faced with one major challenge of security in the form of side channel attacks against highly intensive 3D printing systems. The smartphone formulated Intellectual property (IP) of side channel attacks investigate against 3D printer in the physical domain through reconstructed G-code file through primitive operations. The smartphone (Nexus 5) solved the main problems such as orientation fixing, model accuracy of frame size and validate the feasibility and effectiveness in real case studies against the 3D printer. The 3D printing estimated value reached 20.2 billion of dollars in 2021. The thermal camera is used for exploring the side channel attacks after reconstructing the objects against 3D printers. The researcher analyzed IoT security relevant issues which were avoided in future by enhanced strong security mechanism strategy, encryption, and machine learning-based algorithms, latest technologies, schemes and protocols utilized in an efficient way.*

1. **INTRODUCTION**

The first-time the term "Internet of Things (IoT)" was used was by Kevin Ashton in 1999, the pioneer of British technology [1-6]. According to Kevin Ashton, Internet of Things defines the system of physical objects in the world that connect to the internet via a sensor. Ashton has also invented the term Radio-Frequency Identification (RFID) that tags the physical objects to the internet for the purpose of counting and tracking of goods without any human interference [2]. The IoT components were summarized in an equation form with the combination of sensors, physical objects, controller and actuators [7].

The term "Internet of Things (IoT)" acts as an umbrella keyword that covers the various features such as the extension of the internet, the web as physical realm, deployment of extensive embedded distributed devices, sending and the actuation abilities [8]. The term IoT is also called future internet [9].

The description of the Internet of Things is related to different definitions used by several groups for promoting the particular concept in the whole world. According to the Internet Architecture Board's (IAB) definition, IoT is the networking of smart objects, meaning a huge number of devices intelligently communicating in the presence of internet protocol that cannot be directly operated by human beings but exist as components in buildings, vehicles or the environment. According to the Internet Engineering Task Force (IETF) organization's definition, IoT is the networking of smart objects in which smart objects have some constraints such as limited bandwidth, power and processing accessibility for achieving the interoperability among smart objects. According to the IEEE Communications category magazine's definition, IoT is a framework of all things that have a representation in the presence of the internet in such a way that new applications and services enable the interaction in the physical and virtual world in the form of Machine-to-Machine (M2M) communication in the cloud. According to Oxford dictionary's definition, IoT is the interaction of everyday object's computing devices through the Internet that enables the sending and receiving of useful data [2, 10].

The term Internet of Things (IoT) according to the 2020 conceptual framework is expressed through a simple formula such as: -

$$IoT= Services+ Data+ Networks + Sensors \quad [11]$$

The Internet of Things deals with billions of human and computers to learn about the things such as sensors, actuators and services, and are interconnected with the objects [3, 12].

The IoT is the heterogeneous type of physical devices and communication with the help of unique identifiers in the presence of the internet [13]. Simply, the term IoT is a worldwide network of intelligent objects that are interconnected and uniquely representable on the basis of communication-based protocols [3]. The ubiquitous computing technology related trend improve from the latest technology in the form of the Internet of Things. The Internet of things' objects enable the connection with internet to work anytime, anywhere and anyplace [3, 14].

The IoT four key technological enablers are: -

- ❖ For tagging the things RFID technology used
- ❖ For sensing the things sensor technology used
- ❖ For thinking the things smart technology used
- ❖ For shrinking the things Nanotechnology used [14]

The Internet of Things as a conceptual point of view is majorly built from three pillars relevant to interconnection of smart objects, including identification, communication, and interaction of anything. These three main pillars of interconnected objects are considered as system level characteristics of IoT. According to the system level perspective, IoT act as a vastly dynamic and distributed network system which is composed of smart objects making and consuming information. According to the service level view, IoT represents the integrated intelligent objects' functionality. According to user perception, IoT behaves as a huge amount of responsive services that support and answer the ordinary activities of end-user related activities this is because humans depended upon M2M communication-based devices [8].

The IoT is very famous for the greater enhancement of the latest field of ubiquitous computing, Wireless Sensor Network (WSN) and Machine-to-Machine (M2M) based communication. The IoT vision offers the high-level opportunities of multiple users, manufacturing, and industries. The Internet of Things vision basically requires the advancement of realization process for technologies and applications expanding in Information and Communication Technologies (ICT) fields [11, 13].

The IoT global market, connected with heterogeneous different category devices, increased up to 28 billion approximately in 2020. The IoT is a first-of-kind award winning open source platform that provides discovery and integration of IoT-based devices [15].

The IoT is a fusion of heterogeneous networks including chip technology that scopes gradually more and more, expanding due to the rapid growth of internet applications such as logistics, agriculture, smart community, intelligent transposition, control and tracking systems. According to researcher's analysis, in 2020 IoT objects will be semi-intelligent and an important part of human social life [16].

The IoT visions cover three main aspects consisting of internet vision (smart intelligent object interconnection), things vision (sensors and RFID play vital role for tracking anything) and semantic oriented vision (sensors collect a huge amount of data and retain the useful information) [17].

The IoT devices have border vision which covers several services, consisting of earthquake monitoring systems, building health system, landslides detection, energy management( smart parking and lighting), automation of public building and air quality of noise monitoring system [18].

## 2. ARCHITECTURE

The Internet of Things is considered as the third wave of the World Wide Web (WWW) after static web pages and social networking's based web. The IoT is a worldwide network that connects different type of objects at anytime and anywhere via a popular internet protocol named Internet Protocol(IP) [13, 19].

According to most of the researcher's opinions about conventional IoT architecture, it is considered as three layers: -
- ❖ Perception Layer
- ❖ Network Layer
- ❖ Application Layer

In other aspects, some researchers analyzed one more layer which is also included in IoT's latest architecture that is a support layer that lies between the application layer and network layer. The support layer consists of fog computing and cloud computing. The cloud computing is also the hottest topic today in research. The mostly finalized researchers' view about IoT architecture is shown in Figure 1, in given below [12, 13, 16, 18, 20-22].

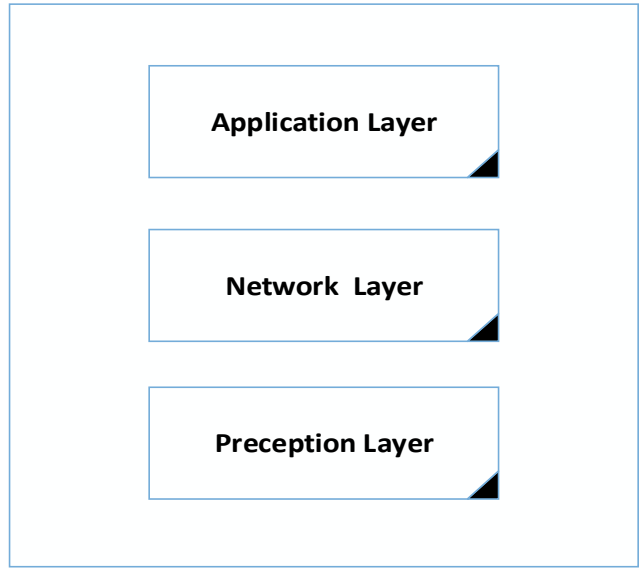

*Figure 1    IoT Architecture [12, 13, 16, 18, 20, 21]*

The above Figure 2.1. represent the basic architecture of IoT. But some of the researchers think that IoT consists of four layers of architecture. The fourth layer is considered a support layer (Technologies used in this new layer are cloud computing, intelligent computing, Fog computing etc.) that lie between the perception and network layer of IoT conventional architecture [4, 20, 23].

The perception layer is also called the recognition layer [20]. The perception layer is the lowest layer of the conventional architecture of IoT. This layer's main responsibility is to collect useful information/data from things or the environment (such as WSN, heterogeneous devices, sensors type real world objects, humidity, and temperature etc.) and transform them in a digital setup. The main purpose of objects is unique address identification and communication between short-range technologies such as RFID, Bluetooth, Near-Field Communication (NFC), 6LoWPAN (Low Power Personal Area Network) [13].

This layer is the brain of conventional IoT architecture [13]. This layer's main responsibility is to help and secure data transmission between the application and perception layer of IoT architecture [20]. This layer mainly collects information and delivers to the perception layer toward several applications and servers. Basically, this layer is a convergence of internet and communication-based networks. According to a current study performed on several communication-based technologies, researchers concluded that the network layer is the most developed layer of conventional IoT architecture. It is the core layer (network layer) of IoT that is capable of advancing the information for relevant procedures. The data processing relevant tasks handled IoT management. This layer also ensures unique addressing and routing abilities to the unified integration of uncountable devices in a single cooperative network. Various types of technologies are contributed for this phenomenon such as wired, wireless and satellite. The implementation of 6LoWPAN protocol towards IPV6 for unique addressing of devices IETF demonstrates a high degree of effort involved [13].

The application layer is considered as a top layer of conventional IoT architecture. This layer provides the personalized based services according to user relevant needs [20]. This layer's main responsibility is to link the major gap between the users and applications. This IoT layer combines the industry to attain the high-level intelligent applications type solutions such as the disaster monitoring, health monitoring, transposition, fortune, medical and ecological environment, and handled global management relevant to all intelligent type applications. According to latest researchers' opinions about IoT architecture, these are five layers as shown in Figure 2.2 [13].

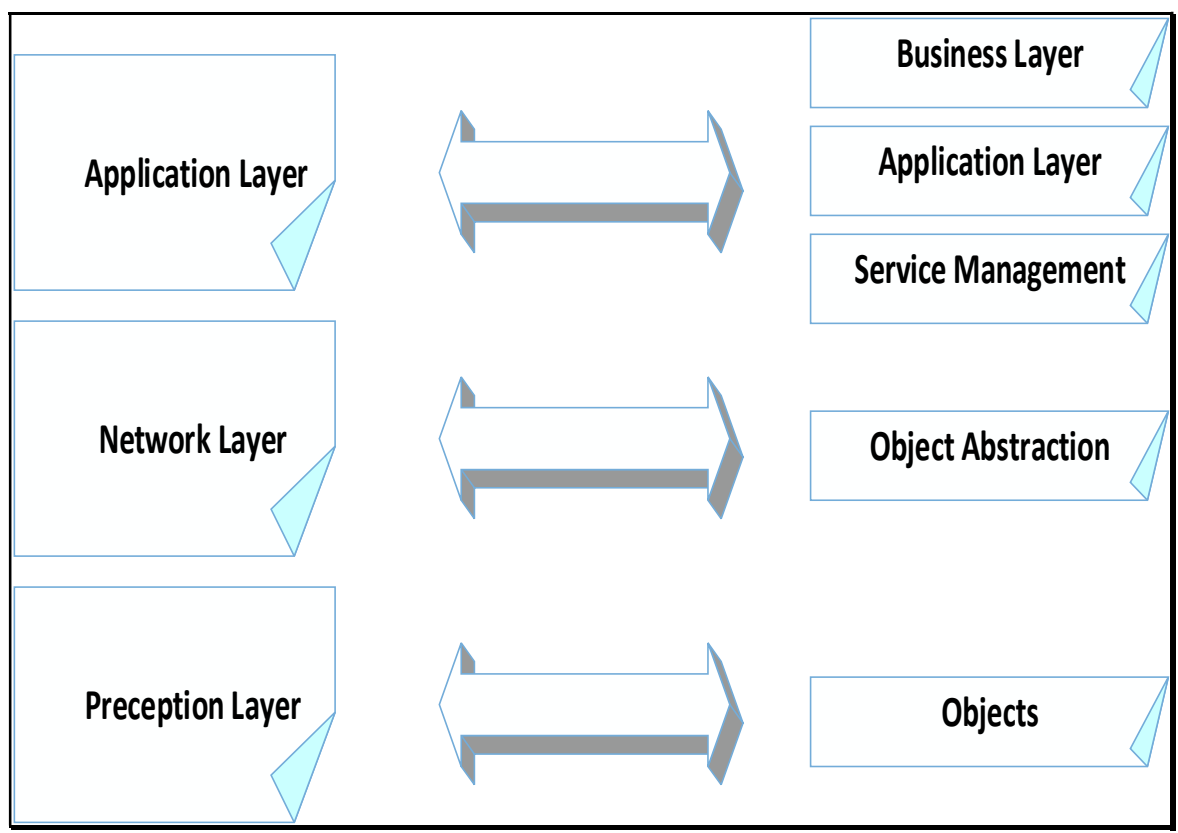

*Figure 2    Enhanced Five Layer IoT architecture [13]*

The above Figure 2.2 shows the five layers that IoT architecture represents. The bottom layer of IoT architecture perception layer represents an object layer. The object layer's main responsibility is to collect data from different heterogeneous category devices and then process and digitize the data. It also transfers the processed data into upper layers of IoT architecture. The middle layer of conventional IoT architecture network layer represents an object abstraction layer. The object abstraction layer acts as a mediating layer between service management and the object layer. In object abstraction, RFID, WIFI and Third Generation (3G) communication technologies are used. The upper layer of IoT architecture, the application layer, is further divided into three sub-layers due to different functionalities. The service management layer's main responsibilities are facilitating information processing, decision-making, and control of pairing requestor information processing for relevant tasks. The application layer provides the customers with smart high-quality facilities according to the pre-request of the customers. The Business layer represents the business model and data that's been received from the application layer [13].

The IoT awareness depends on the compulsory components. According to the high -level functionality of IoT, components are characterized in such a way:

- ❖ Hardware
- ❖ Middleware
- ❖ Presentation

Thus, few additional taxonomies are obtainable for characterization of IoT components, such as: -

- ❖ Data acquisition
- ❖ Communication
- ❖ Computation
- ❖ Services
- ❖ Visualization [3, 8]

The data acquisition with respect to IoT represents a necessity to collect data from different types of objects/devices and share them with multiple devices and IoT applications. The efficient and reliable data aggregation methods basically increase the lifetime of the network by using a sensor [3].

It is facilitated in several technologies that consist of RFID, camera, Barcodes, sensors, actuators, GPS (Global Positioning System) terminal etc. In the above-mentioned technologies, RFID and sensors offer a wide-range of benefits over the data gathering methods. The short-range communication-based technologies facilitate useful information sharing in a different category of heterogeneous devices within the IoT environment [13].

The IoT devices majorly cover a wide variety of applications all over the world, such as healthcare, transposition, agriculture, industry, market, smart home, smart school, smart city and vehicles etc [24, 25].

## 3. TECHNOLOGIES

The Internet of Things (IoT) components basically cover a variety of the latest technologies used in the present era. Some of the major technologies are described below.

The RFID is integrating a Radio Frequency (RF) technology via electromagnetic spectrum to identify unique objects or devices. RFID is similar to a barcode reader based technology but the performance is more efficient [3, 13]. The RFID technology consists of an antenna, transponder, and transceiver. The RF signal is transmitted through an antenna that activates the transponder and data triggers perform the desired action with the help of a programmable logical controller. After all RFID tags the objects that are readable at a certain distance in an efficient way. In RFID technology tags, read collision can occur due to one of the technical issue that is managed via intelligent sensors such as microelectromechanical systems (MEMS). However, RFID range is approximately 3-10 m. RFID is used in indoor moving navigation, smart parking, and battery-less remote control devices applications in the one-way communication-based model [13]. The RFID is a useful IoT architecture layer of the perception layer for object identifying, tracking and exchanging useful information [21].

NFC is a set of the communication-based protocol used for communication between two devices in the range of approximately 10 cm. In this two-way short-range communication, one of the devices must be portable for the purpose of finding a suitable location. The Full-NFC qualified devices container read useful information and stores them in passive NFC-based tags, exchanging information between them and behaving as a smart card for all identification and transactions. The most used applications are smartphones, parking meters, E-ticket booking etc.

Bluetooth is also a short-range communication technology useful for communication between two devices up to almost 100 m in range. It works similarly to the master-slave based architecture. It is designed for low consumption of power. The most common used applications are home automation, communication with peripherals etc.

The collected data from data acquisition components will be transferred to the network. The network will be consumed and processed via different applications. The network will be supported by a variety of communication-based technologies such as RFID, WIFI (Wireless Fidelity), WiMax (Worldwide Interoperability for Microwave Access), NFC, xDSL (x Digital Subscriber Line), Bluetooth, Ethernet, PLC (Power Line Communication) and cellular based networks. The above-mentioned technologies such as xDSL and Ethernet create a higher date rate due to the lie in wire-based technology. Many others mentioned that technologies with lower data rate tend to have the higher flexibility that lies in wireless-based technology.

The xDSL communication technology acts in wired medium. In wired medium copper cables or a twisted pair of cables used that coverage the approximately 1.3 Km range. The XDSL technology main limitation is asymmetric communication.

The Ethernet technology also acts in wired medium. In this technology, copper cables are used to cover the approximate range of 50-70 Km. The Ethernet main limitation is physical medium.

The WIFI is an IEEE 802.11 standard that uses basic radio waves for communication of different devices. Basically, it is a wireless network based technology that provides coverage approximately 100 m in range. One of the biggest limitation is interference facing within WIFI communication.

WiMax is also a wireless network based technology that provides the coverage 50-70 Km in range. The main limitation of this technology is the low data rate in the real world with a sensitivity to weather condition.

The cellular is a commonly used wireless based technology that provides an approximate coverage of 10 m -100 km in range. One of the biggest limitations is restricted wireless spectrum utilized.

The PLC is also an important technology that creates the network through electrical power based system. The PLC is cost effective technology compared to wired technology. It provides coverage approximately 15oo in premises and 100 m between devices. The main limitation is mutual interference with other technologies.

The IoT is linking with a diversity of data source that generate a large amount of data. The IoT is basically considered a computational grid that consists of various type of devices and software based applications that are capable of processing the above generating a large amount of data. In IoT computation different hardware based components and platforms are used such as raspberry, Panstamp, Arduino, TinyOS, RiotOS, LiteOS and vital software platforms. The IoT computations and analysis are varied across a large area from agriculture to health applications. Furthermore, researchers have identified an additional area of computation for working to enhance the communication of IoT.

The IoT services are simply categorized in four main sections such as: -

- ❖ Services relevant to identity (either Active or Passive)
- ❖ Services relevant to information aggregation
- ❖ Services relevant to collaborative-aware
- ❖ Services relevant to ubiquitous devices

The identity services have an additional two major components such as namely identifiers and read devices. These types of services work by first reading the device, reading the identifiers and then sending the encoded information to the server. At the same time, information aggregation requires data from sensors that process, transmit and report IoT relevant applications. The next stage collaborative-aware using this data for decision making. The last ubiquitous provides the services to anyone, anywhere at any time [13].

The visualizations are vital components of IoT-based applications which allow user interaction with the environment [3].

The visualizations provide interaction between IoT objects, the user and a suitable environment.  It also holds two main things such as events detection and visualization of raw and modeled data for user preference [13].

The latest smartphone based technology is utilized against 3D printers. The 3D printing is also called Additive Manufacturing (AM) [26].

 The AM is a layer by layer process of fused materials in Cyber-Physical System (CPS) to produce 3D objects [26-28].

The 3D printing has been increasingly used for fabricating sensitive products in the industry field. The 3D printers have three main advantages: -

- ❖ Efficiency (Production is cost-efficient and fast)

- ❖ Creativity (More flexible in complex based geometric construction)
- ❖ Accessibility (Affordable price point of view)

The global market for 3D printing is estimated approximately 20.2 billion of dollars in the next coming years of 2021 onwards. The Intellectual Property (IP) of sensitive products has been given particular attention with respect to security over last two years. 3D printing is basically divided into cyber and physical domains. But IP protection issue in the physical domain still not explored yet [27].

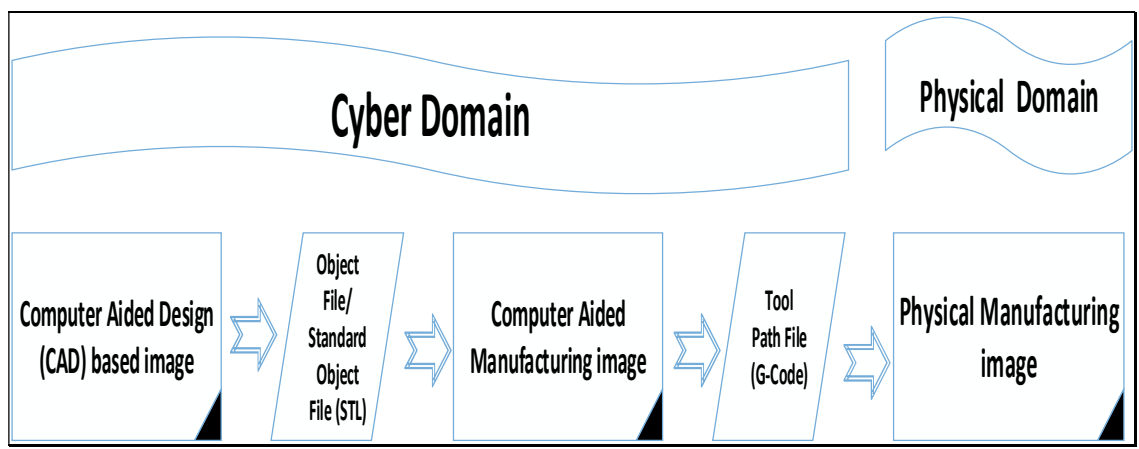

*Figure 3   3D Printing chain includes Cyber and Physical domain [27]*

The above Figure 3 represents a standard 3D printing chain in the cyber and physical system. The first step of the 3D printing chain is creating object models in Computer Aided Design (CAD) and converting CAD models in Standard Object File (STL). The second step of the 3D printing chain is after receiving STL file slices, the models in a uniform layer form in Computer Aided Manufacturing (CAM) and generate the tool path file in the form of G-Code. The G-code normally contains information of 3D digital designs. These two steps are included in the Cyber domain. The third step 3D printers physically manufacture and fabricate the image in the physical domain. The cost effectively Fused Deposition Modeling (FDM) type technology is used in 3D printers [27].

The 3D printer components consist of a fan, extruder and steeper motor, which perform all primitive operations in the physical domain by providing the information from the cyber domain (G-code) [28].

# 4. Literature Work

The security is an essential part of the rapid development of IoT-based applications environment in the present era. The traditional IoT security architecture and standard of IoT is not enough for the user using different intelligent devices. The dynamic defense based mechanism is required for IoT device security issue, which are being solved in better and more efficient ways such as how the Computer Aided Design (CAD) technique are adopting a low-cost solution compared to expensive hardware strategy utilized [13]. The main security goals of IoT devices are Confidentiality Integrity and Availability (CIA) [4, 21]. The novel techniques are used for end-to-end security and privacy managed for next generation IoT-based systems that have a significant impact with respect to performance evaluation techniques [15].

The major IoT security challenges are secured healthcare system and transposition system for saving human life and also preventing financial loss. The IoT architecture layers all faced security issues in term of security attacks. The security requirements are different according to different applications. The IoT architectures' top layer of the application layers' main security issues til today are data sharing that protect user privacy and access controls. The other attacks also faced on application layers include phishing, malware and X scripts. The IoT architecture core network layer faced main security challenges like integrity and confidential data. The other problems on the network layer are Denial of Services (DoS), eavesdropping, a man in the middle, heterogeneity, RFID interference, Node jamming in the WSN and network congestion attack. The IoT architectures' lowest layer, the perception layer, faces the major security challenge of cyber-attack. The other attacks faced in the perception layer are a fake node, malicious code injection, protection of sensor data, side channel attack. These attacks destroy any type of applications in IoT architecture if proper security mechanism, algorithms, and technologies are not implemented in time. The researchers analyzed the IoT architecture layers high impact on security attacks are node tempering, fake node, malicious code based injection, heterogeneity type of network problem, data access and authentication problem, malicious X scripts and malware type attacks [21, 23].

The IoT architecture security requirements are different at different layers. Researchers analyzed that the application layer security requirements are authentication and key management, privacy protection, security education and management. The network layer security requirements are identity authentication, encryption mechanism, and communication security mainly. The perception layer security requirements are lightweight encryption technology, protection sensor data, and key management. The IoT security challenges involved different encryption mechanisms such as end-to-end encryption and hop encryption. The IoT security challenges involved different communication protocols such as Transport Layer Security (TLS), Secure Socket Layer (SSL), Internet Protocol Security (IPSec). The IoT security challenges are controlled through the implement action of cryptographic algorithms such Advance Encryption Standard (AES) for confidentiality, Rivest Shamir Adelman (RSA) for the digital signature of the key, Diffi-Hellman (DH) for key agreement and Secure Hash Algorithm (SHA) for integrity [20].

The IoT security is dependent on three things data confidentiality, privacy, and trust. The IoT security goals are achieved in a better way if the above mentioned three things are utilized for all of the users of IoT efficient and reliable way. The three IoT security challenges are graphically represented in Figure 4 [8].

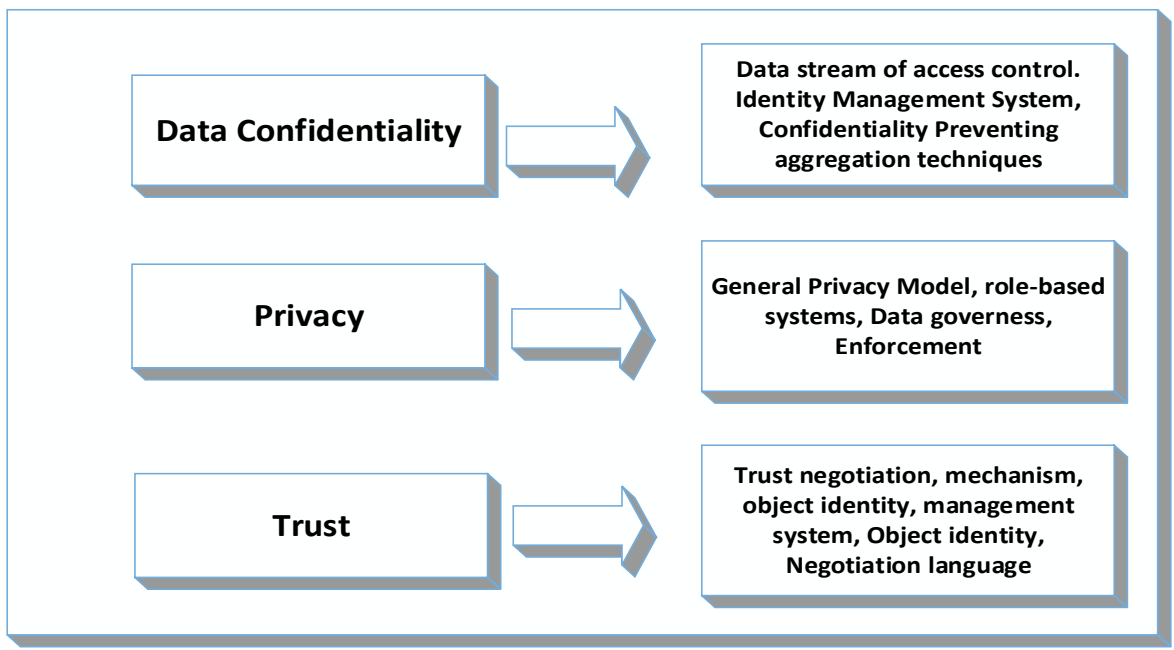

*Figure 4   IoT Security Challenges [8]*

The smartphone-based side channel attacks explored in the physical domain against 3D printers [26, 27]. The researchers analyzed 3D printing mechanism and exploring the side channels (Magnetic and Acoustic side channel attack) via smartphone, and validate the feasibility and effectiveness in a real case study against the 3D printers. The researchers also solved the model accuracy of frame size in 200 ms (time), evaluation and smartphone (Nexus 5) orientation fixing the related problem in the physical domain against 3D printers [27]. The researchers analyzed the relationship between the Cyber domain (G-code) and the physical domains' relevant data in additive manufacturing [26].

The attackers are so intelligent, using audio voice via smartphone that were closest to 3D printers and recorded sounds while printing objects for attacking 3D systems easily. So, according to the attackers' minds researchers divided the printing mechanism into two phases. The 1$^{st}$ phase is the training phase and 2$^{nd}$ phase is the attacker phase. In the training phase, the audio signal is recorded, pre- processed and extracted to map with G-Code. The next G-code mapped features were supplied to regression based model that classified for training their learning based algorithms. This model was also used for predicting the speed of nozzle [26].

The InfraRed Thermography (IRT) developed the concept of thermal imaging. The thermographic camera normally defects infrared radiations via the electromagnetic spectrum (9 – 14 µm range approximately) and produces thermogram based images. One of the major benefits of thermography technology is that it is used in the military to allow clear visibility in day and night human and warm-blooded animals. The additive manufacturing is a layer by layer process of deposition of the filament melted nozzle through heat, and this heated nozzle is captured by a thermal camera that then distinguishes between nozzles and others objects in any environment. The 3D attack model reconstructs objects through tracking the movement of nozzle [28].

The network printers gained attention in regards to security in 1996 when PostScript file of I/O primitive was highlighted by Silbert. In this paper security attacks were analyzed regarding printers' Printer Exploitation Toolkit (PERT). This involved evaluating 20 printers through PERT for vulnerability to different attacks. Because no efficient tool exists for evaluating security of printer. The printer attacks such as Denial-of-Service (DoS) damage the documents processing and physically damage Non-Volatile Access Memory (NVRAM) [29].

The side channel attacks gained attention in 2007 when mobile devices were destroyed via this type of attack. The side channel attacks affected both security and privacy relevant sensitive information of user mobile devices. The Side Channel attacks are systematically categorized in two axes of orthogonal.

- ❖ Active vs Passive (depended upon attackers actively influence the performance of devices or passively leaking based information observe).
- ❖ Invasive vs non-invasive vs semi-invasive (Depended upon the attacker removes passivation layer of the chip, chip depackages or packaging manipulation not occur).

The side channel attacks leakage of information is categorized in two forms. The first unattained information leakage (Including execution time, power consumption and EM emanation)and second information publishers' purpose (Including footprint, sensor information, and data consumption) [30].

The IoT-based latest technology has an ability to integrate networking capabilities of WSN and smartphone devices in order to achieve monitoring objects, tracking the mobility of devices. The monitoring and mobility of each devices' movement monitor and tracking facility are provided to the user at home. The smartphone-based sensors attained data in 3D axis consists of X, Y, and Z. In this paper, smartphone (IPhone's) sensor based environment is observed [31].

## 5. Evaluation Methods

The evaluation methods section consists of previous research methodologies experimental result and various evaluation based techniques relevant to discussion in detail.

The Fused Deposition Modelling (FDM) most commonly cost-effective technology is used in the 3D Printing process. In this proposed solution two algorithms in G-code are implemented. The first algorithm reconstructed the G-code for 3D printing primitive operation layers for exploring the magnetic and acoustic attack. The 2nd algorithm was used to smooth the layers. The researchers solved the major challenge of model accuracy for getting better performance by increasing the frame size. The authors defended the 3D design mechanism from side channel attacks with the help of two defense methods based on the use of hardware and software. In software, the defense method author protects the 3D printer with dynamic path strategy and dummy task injection. In the hardware defense mechanism authors defend the 3D printer from side channel attack by shielding the hardware and generating more side channel interference for attackers.

The researchers evaluated the existing primitive models and proposed a new model for analyzing performance in a real case study. The FDM type 3d printer (Ultimaker 2) and Smartphone (Nexus 5) is used for exploring the side channel signals. In this smartphone is a built-in application for data, recording is used to collect the side signal information. Both printer and smartphone built in sensors have owned their coordinate and configuration. The audio and magnetic data is recorded at different frequencies. According to the aim of the proposed model nozzle printing speed is set 180 mm./min and alignment speed is set 7800 mm/min for getting a high quality of printing product. In primitive operation models the side channel data is a filter on Savitkzy-Golay and segment the signals in separate fixed size frames of 200 ms. The authors partition fixed frames into training and testing set at different models (Layer, Head, Axial and Directional movements). The frame size is an important factor directly affected by the model's performance. The performance of models gradually increased by increasing the frame size. The larger frame size means each frame has more characteristics' information contained that deduced the high dimensional feature domain.

This technique is evaluated in real case scenario by selecting the rectangle first as a regular shape and involving all primitive operations. The purpose achieved with G-Code file is a generation of four layers of objects in a 90mm*90mm rectangle including 1 mm height. In each layer, the reconstructed shape is fit to the original triangle. The Y-directional model performed the better result as compared to X-directional movement. For evaluation, the reconstruction performance error metric introduces in 3D printing attack. This error metric eliminated the traditional errors metric with Mean Tendency Error (MTE) for accessing the geometrical reconstruction based on relative shape difference. The lower MTE over different layers indicates the attack method can accurately and robustly reconstruct the original shape. The average MTE of four layer objects is 5.87 % and real complex design of 10 layers' object (90mm*45mm with 1 mm height) is 9.67 %. The variations between the layer's smooth layer algorithm are performed and adjust the contour outliers. The authors evaluate the smartphone orientation as independent at different angles (0, 30, 60, 90) degrees [27].

The learning based algorithm is used before using model attacks. The trained algorithm data consist of G-code that move data at a different speed (500 to 4800 mm / minutes) in X, Y or X, Y, Z directions. The audio signals are recorded via training learning based algorithms, with a total length of 1 hour and 48 minutes and a frequency extract of 70 Hz to 10 KHz for 3D printers. The accuracy of regression algorithms is measured in terms of Means Square Error (MSE) that normalized data with zero means unit variance. The MSE has a relatively higher value as compared to the regression model. The regression model motion Y axis is higher MSE value as others axes. The researchers analyzed that the 3D printer's nozzle movement changes direction with the distance of successive features of frames. The researchers tested the accuracy of the attack model in terms of speed of printing, the dimension of objects and complexity of objects. The acoustic side channel attack's average accuracy of axis prediction is 66.29 % and the average length prediction error is 20.91 %. The post proceeding average axis prediction is 78.35 % and pre-diction error is 17.82 %. The attack model accuracy axis prediction accuracy is 92.54 % and length prediction error is 6.35 % achieved. The tested model highlighted serious physical to cyber domain acoustic side channel attacks against 3D printers [26].

The researchers evaluated 20 network printers from different manufacturers After installing the latest firmware of printers three printers broke before starting the evaluations of printer attacks. The DoS attacks are applicable to all printers with few line of codes. The attackers easily disclosed the printing information such as memory access, file system access, print job capture and credential disclosure. Beyond the cloud computing technology researcher analyzed 3D printers' growing trends. This study provides IoT devices' security attacks relevant awareness [29].

The side channel attacks which destroyed the smartphone sensitive data, consist of: -
- ❖ Local Side-Channel Attack (Protection cryptographic implementation and protecting user sensitive information)
- ❖ Vicinity Side-Channel Attack (Preventing network based traffic analysis)
- ❖ Remote side-Channel Attacks (Permission, Keyboard layout randomization sampling frequency, noise injection and preventing microarchitectural attacks)

The side Channel attacks on smartphone can be avoided if an authentic code analysis tool is used for different types of applications such as google play store [30].

## 6.    Conclusion

Therefore, it is concluded that the IoT architecture security related issues facing all layers today are an active topic in research industry and academia. The security issues in the IoT application layer are data access and authentication, phishing attacks, malware attacks, malicious Active X scripts. The security issues in the IoT network layer are network congestions, RFID interference, eavesdropping attack, routing attack, denial of service, Sybil attack node jumping in WSN and heterogeneity problem. The security issues in the IoT perception layer are a fake node, malicious code injection, protecting sensor data, mass code authentication, physical damage and node tempering issues [23]. Hence, it is also concluded that newly integrated security authentication mechanisms, technologies, and algorithms can be used for reliable and secure M2M communication [13, 16, 20, 23]. Hence, it is also resolved that the useful data actuation issues through compressed sensing framework saved energy, communication-based resources in information and network system efficiently. It is concluded that IoT development of vision embedded devices that merged virtual and real world, and opened new doors in the fields of research and business [8, 17]. Therefore, it is also concluded that smartphone based acoustic side channel attack cannot be explored properly in the physical domain in Sensitive Intellectual Property (IP) against 3D printers [26-28].

## 7. Future Findings

The IoT is a very interesting and famous research field in the present era. But IoT field has required solving a variety of research questions at different architecture layers and other aspects such as one of major information security. The following are some of the latest research challenges: -

- ❖ The researchers identified that the major research challenges faced currently among the IoT field are security, reliability, mobility, scalability, interoperability, physical domain security, authentication, risk assessment, intrusion detection based techniques and performance issues [4, 13].
- ❖ The researchers recognized several open-handed challenges regarding IoT, such as security issues, by exchanging information within the IoT [13, 20, 23, 32].
- ❖ The researchers identified the numerous other challenges regarding IoT, such as sensing the complex nature of the environment, several connectivity choices, privacy protection and growing the architecture.
- ❖ The researchers identified one of the huge challenges faced today in IoT as storing objects' unique addressing and representation of exchanged information.
- ❖ The researchers analyzed that the performance of IoT's evaluation gap has not been filled till today [13].

- ❖ The researchers practically analyzed authentication of IoT devices with respect to real world situation or M2M communication devices which are still not solved today [13, 23].
- ❖ The extraction of useful information in complex based sensing environment at a variety of three-dimensional and sequential resolution is still not solved in Artificial Intelligence (AI).
- ❖ The evaluation of next generation mobile applications security issues still not solved yet [3].
- ❖ The authentication of IoT devices in world scenario is still not solved yet [13].
- ❖ The researchers analyzed the most important application for human life point of view healthcare system and transposition system security issues are still not solved yet [23].
- ❖ The limitation of this paper is distance effect, printing speed effect (Delay more), position effect, ambient nice effect, carry on side channel attack and advance shape effects [27].
- ❖ The researcher analyzed physical to cyber domain attacks against Intellectual Property (IP) of sensitive products against 3D printers cannot explore properly still yet [26, 27].
- ❖ The researchers analyzed the integration of IoT-based devices and the initial phase of cloud computing cannot be extended all type of application domain sites due to inadequate security architecture [5].
- ❖ The integrate adaptive frame size increased the accuracy of attack models [26].
- ❖ The IoT-based appliances on smartphone face main threat risk at physical possessions [30].
- ❖ The extensive design of IoT-based devices has faced several issues such as deployment, mobility, cost aspect, heterogeneity, communication modality, coverage, connectivity, infrastructure, network size, network topology, Quality of Service (QoS) requirements, lifetime etc. [6].